\documentclass[12pt]{article}
\usepackage{amsmath}
\usepackage{epsfig}

\begin{document}

\date{\today\\ \vspace{1cm} }

\centerline{\large \bf  One Special Identity between the complete
elliptic integrals}

\vskip .15in

\centerline{\large \bf of the first and the third kind}

\vskip .4in

\centerline{Yu Jia }

\vskip .2 in

\centerline{Institute of High Energy Physics, Chinese Academy of
Sciences}

\centerline{Beijing 100049, China}
\vskip .7 in

\centerline{\bf Abstract} \vskip .15 in

I prove an identity between the first kind and the third kind
complete elliptic integrals with the following form:
\begin{eqnarray}
& & \Pi\left({(1+x) (1-3 x)\over (1-x) (1+3 x)}, {(1+x)^3 (1-3
x)\over (1-x)^3 (1+3 x)} \right) - {1+ 3 x \over 6 x}\,K \left(
{(1+x)^3 (1-3x)\over (1-x)^3 (1+3x)}\right) \nonumber\\
&=&
\begin{cases}
 0, \hspace{2.5 in} (0< x < 1) \nonumber\\
-{\pi\over 12} {(x-1)^{3/2} \sqrt{1+3 x}  \over x }. \hspace{1.45
in}(x<0\;\;{\rm or}\;\;x>1)  \nonumber
 \end{cases}
\end{eqnarray}
This relation can be applied to eliminate the complete elliptic
integral of the third kind from the analytic solutions of the
imaginary part of two-loop sunset diagrams in the equal mass case.

The validity of this relation in the complex domain is also
briefly discussed.

\vskip .4in


\newpage

Elliptic integrals, as one of the most important classes of
nonelementary functions,  have found numerous applications in many
branches of engineering and physics. Having been comprehensively
studied by many eminent mathematicians over centuries, a vast
amount of knowledge on the elliptic integrals, as well as their
close cousin, elliptic functions, has been
accumulated~\cite{Byrd:Friedman:elliptic:handbook,
Abramowitz:Stegun,Gradshteyn:Ryzhik,Mathematica5.0}.

The simplest and particularly important class of elliptic
integrals is the {\it complete} elliptic integrals, totally of
three kinds. The complete elliptic integral of the third kind,
$\Pi$, being the most complicated one,  can be expressed in terms
of the complete elliptic integral of the first kind, $K$, plus
elementary functions and  Heuman's Lambda function or Jacobi's
Zeta function. The last two transcendental functions can in turn
be expressible from the {\it incomplete} elliptic integrals of the
first and second kinds.

The aim of this note is to establish a simple relation between the
complete elliptic integrals of the first and the third kind, for
some specific texture of arguments of course. That is, in this
special case, the complete elliptic integral of the third kind can
be transformed to the complete elliptic integral of the first
kind, plus elementary function at most, without resort to any
other nonelementary function.

There are a variety of conventions adopted in literature in defining
the elliptic integrals.  In this work I find it convenient to use
the same definitions as taken by
\cite{Abramowitz:Stegun,Mathematica5.0} for the first, second and
third kind of complete elliptic integrals:
\begin{eqnarray}
K(m) &\equiv & \int^1_0 d t {1\over \sqrt{(1-t^2)(1- m t^2)}}=
{\pi\over 2} \,   F\left({1\over 2},{1\over 2};1; m \right),
\nonumber \\
E(m) &\equiv & \int^1_0 d t  \sqrt{1-m t^2\over 1- t^2}= {\pi\over
2} \,   F\left(-{1\over 2},{1\over 2};1; m \right),
\\
\Pi(n,m) &\equiv & \int^1_0 d t {1\over (1-n t^2) \sqrt{(1-t^2)(1-m
t^2)}} =  {\pi\over 2} \,   F_1\left({1\over 2};1,{1\over 2};1;n,
m\right), \nonumber
\end{eqnarray}
where $F$ and $F_1$ denote Gaussian hypergeometric function, and
Appell hypergeometric function of two variables, respectively. In
most practical applications,  the parameter $m$ and the
characteristic $n$ are restricted to be less than 1.  However, it
is worth emphasizing when these arguments exceed 1 or even are off
the real axis, these elliptic integrals are still well defined
mathematically,  though become complex valued in general.

The main result states as follows.  For $\forall\, x\in R$, there
exists an identity
\begin{eqnarray}
& & \Pi\left({(1+x) (1-3 x)\over (1-x) (1+3 x)}, {(1+x)^3 (1-3
x)\over (1-x)^3 (1+3 x)} \right) - {1+ 3 x \over 6 x}\,K \left(
{(1+x)^3 (1-3x)\over (1-x)^3 (1+3x}\right) \nonumber\\
&=&
\begin{cases}
 0, \hspace{2.25 in} (0< x < 1)
\\
-{\pi\over 12} {(x-1)^{3/2} \sqrt{1+3 x } \over x }. \hspace{1.2
in}(x<0\;\;{\rm or}\;\;x>1)
\end{cases}
\end{eqnarray}
The $K$ and $\Pi$ functions with this specific arrangement of the
arguments,  odd as it may look,  are not unfamiliar to particle
physicists. In fact these functions have been encountered in the
analytical expressions for the phase space of three equal-mass
particles~\cite{Almgren:1968,Bauberger:1994by,Davydychev:2003cw}.
To be precise, it is worth pointing out that the coefficient of
$\Pi$ function accidentally vanishes in this example. Note the
3-body phase space can be obtained from cutting the simplest
scalar two-loop sunset diagram. For a general sunset diagram with
a nontrivial vertex structure, or if the power of propagators
exceeds than 1, the complete elliptic integral of the third kind
will inevitably arise in evaluating its imaginary part.  Equation
(2) can then be invoked to trade the $\Pi$ function for the $K$
function, thus considerably simplifying the answer.

In the aforementioned physical application,  simple kinematics
enforces $0< x \le {1\over 3}$, for which both the parameter and
characteristic of $K$ and $\Pi$ are positive and less than 1. This
restriction is not necessary for general purpose, therefore it
will be discarded in the following discussion.

There exist some formulas which transform one $\Pi$ function to
another $\Pi$ function plus a $K$
function~\cite{Byrd:Friedman:elliptic:handbook,Abramowitz:Stegun}.
However, the relation designated in (2), which links one $\Pi$
function to one $K$ function only, is rather peculiar. To the best
of my knowledge, this relation cannot be derived by any known
formula, and it also has never been explicitly stated in any
published work.  For this reason, I feel it may be worthwhile to
report it here.

The strategy of the proof is to construct a differential equation
satisfied by the left side of equation (2),  called $y(x)$ in
shorthand.  Employing the well-known differential properties of
complete elliptic integrals~\cite{Byrd:Friedman:elliptic:handbook,
Mathematica5.0}:
\begin{eqnarray}
{d K(m)\over d m} &=& {E-(1-m)K\over 2 m(1-m)},
\nonumber \\
{\partial \Pi(n,m)\over \partial m} &=& {-E+(1-m)\Pi \over 2
(1-m)(n-m)},
\\
{\partial \Pi(n,m)\over \partial n} &=& {n E-(n-m)K+(n^2-m)\Pi \over
2 n (1-n)(n-m)}, \nonumber
\end{eqnarray}
together with the chain rule, I find that $y$ satisfies the
following first-order differential equation:
\begin{eqnarray}
y' &=& y \,{1+2x+3x^2 \over x (x-1)(1+3 x)}. \end{eqnarray}
The magic is that, after the differentiation, the complete elliptic
integral of the second kind cancels, and the elliptic integrals of
the first and the third kind conspire, in a rather peculiar way, to
cluster into the original form.

This is a basic type of linear differential equation, and the
corresponding solution is
\begin{eqnarray}
y(x) &=& C \, {(x-1)^{3/2} \sqrt{1+3 x}  \over x },
\end{eqnarray}
where $C$ is a constant to be determined. Notice the above
solution possesses a pole at $x=0$ and two branch points located
at $x=1$ and $x=-{1\over 3}$,  and in general one should not
expect $C$ will assume a universal value in the entire domain of
$x$, $R$. I shall attempt to fix the value of $C$ region by
region.

First let us consider the case when $x$ belongs to the open
interval $I_0=(0,1)$. Inspecting Eq.(2), it is easy to see
$y(x_0)=0$ for $x_0={1\over 3}\in I_0$. This initial value cannot
be satisfied unless if $C=0$.  Also note the right hand side of
(4) is a continuous function of $x$ in this interval.  Therefore,
by the existence and uniqueness theorem for linear equation (for
instance, see theorem 2.1 in Ref.~\cite{Boyce:DiPrima}), the
differential equation (4) admits the unique solution $y=0$ in this
interval.

Next I turn to the solution for $x\in I_1=(-\infty,-{1\over 3})$.
Examining the left side of Eq.(2), one readily finds
$y(x_1)={\pi\over 3}$ for $x_1=-1\in I_1$. This initial value can
be satisfied only if $C=-{\pi\over 12}$. Thus by the theorem of
existence and uniqueness, the function in (5) with this value of
$C$ constitutes the unique solution in the interval $I_1$.

There still remain two other intervals, $I_2=(-{1\over 3},0)$ and
$I_3=(1,\infty)$ to be investigated. When $x$ resides in $I_2$,
both $K$ and  $\Pi$ functions become complex-valued, and the left
side of (2) turns to be purely imaginary; whereas as $x\in I_3$,
though both $K$ and $\Pi$ become also complex, the left side of
(2) nevertheless  is real. There are no simple initial values can
be inferred in these regions. I performed a numerical check with
the aid of the computing package
\textsc{Mathematica}~\cite{Mathematica5.0}, and find the solution
(5) with $C=-{\pi\over 12}$ in these two regions agree with the
left side of Eq.(2) to the 14 decimal place.

Equation (2) can have some interesting consequences. Here I
illustrate one example:
\begin{eqnarray}
{\Gamma({1\over 4})^2 \over 4\sqrt{\pi}}
&=& K\left({1\over
2}\right) \nonumber \\
&=& {3-\sqrt{6\sqrt{3}-9} \over 2}\, \Pi\left({1-\sqrt{2\sqrt{3}-3}
\over 2},{1\over 2}\right)
\\
&=& {3+\sqrt{6\sqrt{3}-9} \over 2}\, \Pi\left({1+\sqrt{2\sqrt{3}-3}
\over 2},{1\over 2}\right)- \pi \sqrt{2+
\sqrt{3}+\sqrt{7+{38\sqrt{3}\over 9}}}.
\nonumber
\end{eqnarray}
The analytic expression for $K({1\over 2})$ in term of $\Gamma$
function is known~\cite{Gradshteyn:Ryzhik}.  However, it is
amusing that these strange looking $\Pi$ functions can also be put
in closed form.

\begin{figure}[htb]
\begin{center}\vspace*{-1cm}
\includegraphics[scale=1.2]{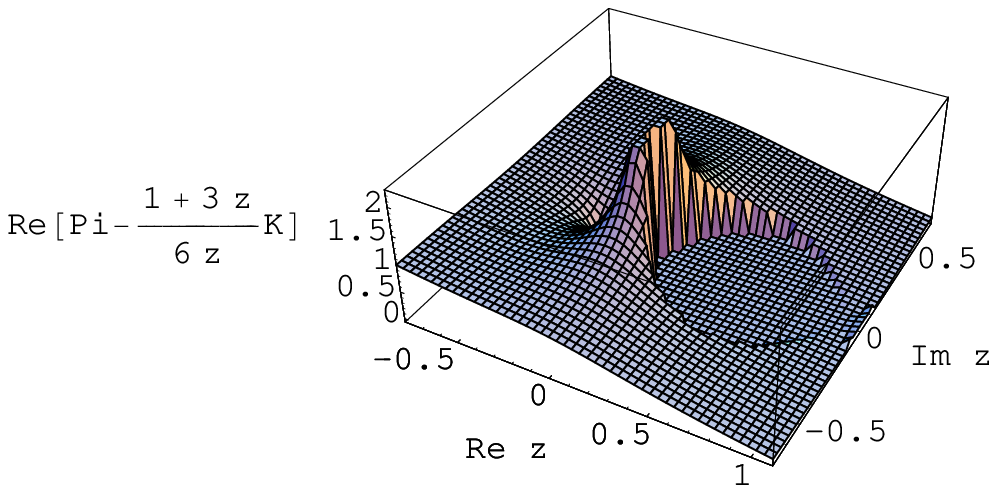}%
\vspace{-1.5 cm}
\includegraphics[scale=1.2]{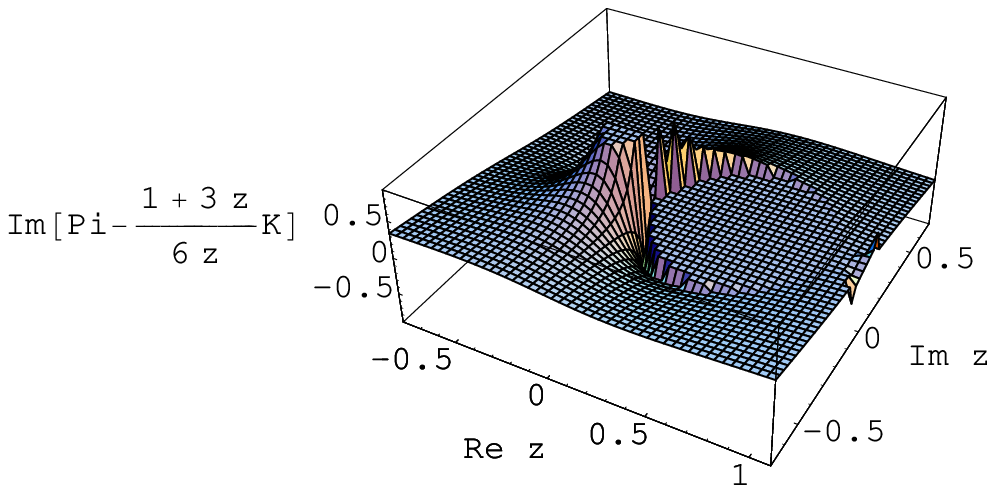}%
\vspace{-1cm} \caption{\label{profile:lhs2} Profile of the left
side of Equation (2) as a complex-valued function, with the real
variable $x$ promoted to a complex variable $z$.}
\end{center}
\end{figure}

\begin{figure}[thb]
\begin{center}\vspace*{-2.cm}
\includegraphics[scale=1.4]{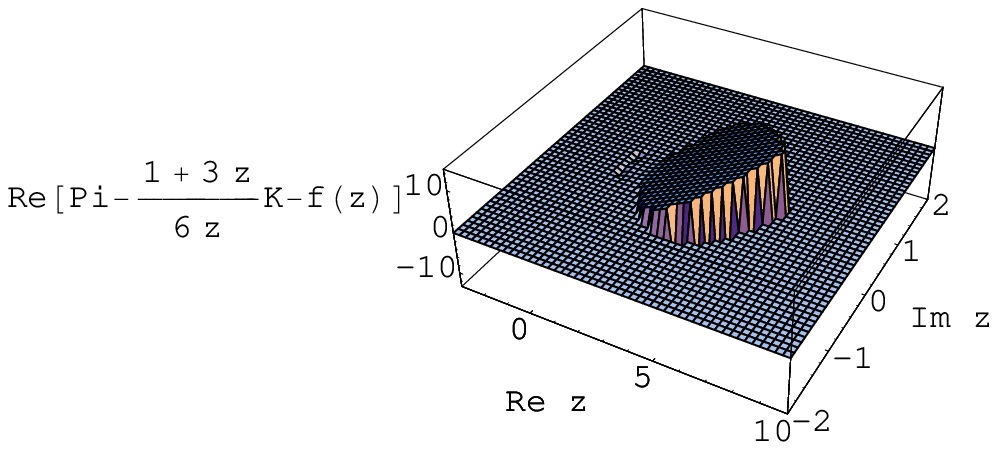}%
\vspace{-2.5 cm}
\includegraphics[scale=1.4]{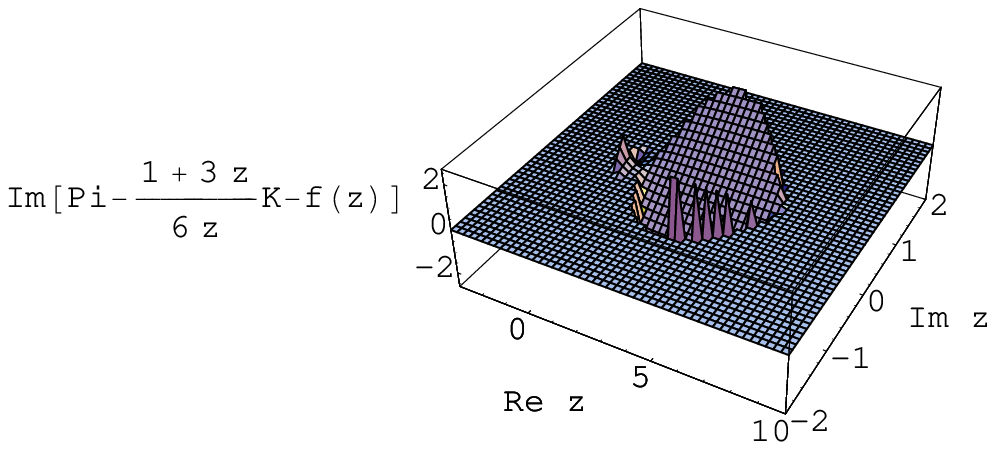}%
\vspace{-2cm}\caption{\label{profile:lhs2:minus:f} Examination of
the validity of second portion of Equation (2) with a complex
variable $z$, where $f(z)\equiv -{\pi\over 12}{(z-1)^{3/2}
\sqrt{1+3 z} \over z }$.}
\end{center}
\end{figure}

Lastly, one natural question may be raised-- how about the
validity of equation (2) when the domain of $x$ is extended from
real to complex?  It is a curious question owning to the rich
analytic structure of elliptic integrals.   Through a numerical
study using \textsc{Mathematica},  I find this relation still
holds in most regions of complex plane. It is most lucid to
demonstrate this examination in plots. The left side of equation
(2) still vanishes in an oval region internally tangent to a
rectangle ($0< {\rm Re}\,z<1$, $|{\rm Im}\, z| < 0.33$), which is
characterized by the basin and plateau in
Figure~\ref{profile:lhs2}. This can be viewed as a generalization
to the first portion of equation (2).  The second part of this
relation is valid almost everywhere else except in a pear shaped
region embedded inside a rectangle ($1< {\rm Re}\,z <6.74$ and
$|{\rm Im}\, z| < 0.97$), which can be clearly visualized in
Figure~\ref{profile:lhs2:minus:f} as the tip of an iceberg
surrounded by the flat, boundless sea.  At this stage I am unable
to provide a rigorous justification for this observation. A
thorough understanding of these features will be definitely
desirable.

\section*{Acknowledgment}

I thank Kexin Cao for  encouraging me to write down this note. This
research is supported in part by National Natural Science Foundation
of China under Grant No.~10605031.



\begin{thebibliography}{20}

\bibitem{Byrd:Friedman:elliptic:handbook}
P.~F.~Byrd and M.~D.~Friedman, {\it Handbook of Elliptic Integrals
for Engineers and Scientists}, Springer Verlag, Berlin (1971).


\bibitem{Abramowitz:Stegun}
M.~Abramowitz and I.~A.~Stegun, {\it Handbook of Mathematical
Functions}, Dover Publications, New York (1972).


\bibitem{Gradshteyn:Ryzhik}
I.~S.~Gradshteyn and I.~M.~Ryzhik, {\it Table of Integrals, Series,
and Products}, Academic Press, San Diego (2000).


\bibitem{Mathematica5.0}
Wolfram Research, Inc., {\it Mathematica Edition: Version 5.0},
Wolfram Research, Inc., Champaign, IL (2003).


\bibitem{Almgren:1968}
B.~Almgren, Arkiv f\"{o}r Physik {\bf 38}, 161 (1968).


\bibitem{Bauberger:1994by}
  S.~Bauberger, F.~A.~Berends, M.~Bohm and M.~Buza,
  Nucl.\ Phys.\  B {\bf 434}, 383 (1995)
  [arXiv:hep-ph/9409388].

\bibitem{Davydychev:2003cw}
  A.~I.~Davydychev and R.~Delbourgo,
  J.\ Phys.\ A  {\bf 37}, 4871 (2004)
  [arXiv:hep-th/0311075].


\bibitem{Boyce:DiPrima}
W.~E.~Boyce and R.~C.~DiPrima, {\it Elementary Differential
Equations and Boundary Value Problems} (2nd Edition), John Wiley
\& Sons , New York (1969).


\end{thebibliography}
\end{document}